\documentclass[preprint2]{aastex}

\shorttitle{Distance and Reddening to the dIrr Galaxy NGC 1156}
\shortauthors{S. C. Kim et al.}

\begin{document}

\newcommand{\hi}{{\sc H~i}\/ }
\newcommand{\hii}{{\sc H~ii}\/ }
\newcommand{\ha}{H$\alpha$\xspace}
\newcommand{\kms}{km~s$^{-1}$}
\newcommand{\Msun}{$M_\odot$} 
\newcommand{\Msunn}{$M_\odot$ }
\def\simlt{\lower.5ex\hbox{$\; \buildrel < \over \sim \;$}}
\def\simgt{\lower.5ex\hbox{$\; \buildrel > \over \sim \;$}}
\def\arcdeg{\hbox{$^\circ$}}
\def\arcmin{\hbox{$^\prime$}}
\def\arcsec{\hbox{$^{\prime\prime}$}}

\title{Distance and Reddening of the Isolated Dwarf Irregular Galaxy NGC 1156}
\author{Sang Chul \textsc{KIM}$^1$, Hong Soo \textsc{PARK}$^2$, Jaemann \textsc{KYEONG}$^1$, 
Joon Hyeop \textsc{LEE}$^1$, Chang Hee \textsc{REE}$^1$, and Minjin \textsc{KIM}$^{1,3,4}$}
\affil{$^1$Korea Astronomy \& Space Science Institute, Daejeon 305-348, Korea\\
$^2$ Department of Physics and Astronomy, Seoul National University, Seoul 151-742, Korea\\
$^3$ The Observatories of the Carnegie Institution for Science, 813 Santa Barbara Street, 
  Pasadena, CA 91101, USA \\
$^4$ KASI-Carnegie Fellow}
\email{sckim@kasi.re.kr, hspark@astro.snu.ac.kr, jman@kasi.re.kr, 
jhl@kasi.re.kr, chr@kasi.re.kr, mkim@kasi.re.kr \\ 
{\bf To appear in PASJ 2012 April 25, Vol. 64, Issue 2} }

\begin{abstract}
We present a photometric estimation of the distance and reddening values
  to the dwarf irregular galaxy NGC 1156, which is one of the best targets
  to study the isolated dwarf galaxies in the nearby universe.
We have used the imaging data sets of the {\it Hubble Space Telescope (HST)} 
  Advanced Camera for Surveys (ACS) High Resolution Channel (HRC) of 
  the central region of NGC 1156 ($26\arcsec \times 29\arcsec$)
  available in the $HST$ archive for this study.
From the $(U-B, B-V)$ color-color diagram, we first estimate the total 
  (foreground $+$ internal) reddening toward NGC 1156 of $E(B-V) =0.35 \pm 0.05$ mag, 
  whereas only the foreground reddening was previously known to be
  $E(B-V)=0.16$ mag (Burstein \& Heiles) or 0.24 mag (Schlegel, Finkbeiner, \& Davis).
Based on the brightest stars method,
  selecting the three brightest blue supergiant (BSG) stars with mean $B$ magnitude of
  $\langle B(3B) \rangle = 21.94$ mag and  
  the three brightest red supergiant (RSG) stars with mean $V$ magnitude of 
  $\langle V(3R) \rangle = 22.76$ mag,    
  we derive the distance modulus to NGC 1156 to be $(m-M)_{0,BSG} = 29.55$ mag and
  $(m-M)_{0,RSG} = 29.16$ mag. 
By using weights of 1 and 1.5 for the distance moduli from using
  the BSGs and the RSGs, respectively,
  we finally obtain the weighted mean distance modulus to NGC 1156
  $(m-M)_0 = 29.39 \pm 0.20$ mag ($d = 7.6 \pm 0.7$ Mpc),
  which is in very good agreement with the previous estimates.
Combining the photometry data of this study with those of 
  Karachentsev et al. gives smaller distance to NGC 1156,
  which is discussed together with the limits of the data.
\end{abstract}

\section{INTRODUCTION}

Dwarf galaxies are an important class of galaxies in studying the evolution of galaxies 
  as well as the cosmological evolution of the universe. 
They have much simpler structures than larger/giant galaxies,
  and are prone to be affected by small perturbations.
It is also easy to observe and study the whole systems of dwarf galaxies
  because of their small sizes
  \citep{kim98, kyeong06, cole07, kyeong10}.
Isolated dwarf galaxies are even better targets for studying
  the evolution of the system
  since they are not affected by any environmental effects
  so that any kind of causes and effects reside in the system itself.

NGC 1156 (UGC 2455, Vorontsov-Velyaminov [VV] 531, PGC 11329, KIG 0121,
  IRAS F02567+2502) is an isolated, Magellanic type dwarf irregular (dIrr) galaxy
  with the morphological type of IB(s)m V-VI \citep{sandage94,devaucouleurs91}.
This galaxy has boxy-like shape and bright blue patches
  implying an active star formation stage, though not triggered
  by any external tidal perturbations \citep{karachentsev96}.
NGC 1156 is one of the highly isolated and less disturbed galaxies, 
  and its nearest neighbors are UGC 2684 and UGC 2716, 
  located more than $10\arcdeg$ away from NGC 1156 \citep{karachentsev96,minchin10}.
Studying the star formation rate, mass loss rate, current star formation activity,
  etc. for NGC 1156 is very interesting, because this galaxy is not thought to be
  disturbed/triggered by any nearby objects \citep{hunter04}.
Recently, using the 21 cm line of neutral hydrogen ({\sc H~i}) 
  observed with the Arecibo $L$-band Feed Array,
  \citet{minchin10} found a new small dwarf galaxy dubbed AGES J030039+254656
  at 35\arcmin~ north-northeast of NGC 1156 (80 kpc in projection,
  $\alpha_{J2000.0}= 03^{\rm h} 00^{\rm m} 38.6^{\rm s}$,
  $\delta_{J2000.0}= +25\arcdeg 47\arcmin 02\arcsec$).
This galaxy has an {\sc H~i} flux of $0.114 \pm 0.032$ Jy \kms,
  giving it a neutral hydrogen mass of $(1.63\pm0.46) \times 10^6$ \Msun.
\citet{minchin10} estimated that the star formation rate and {\sc H~i} mass
  of this galaxy are both around three orders of magnitude lower than 
  in the case of NGC 1156
  and concluded that it is unlikely that AGES J030039+254656, 
  in its current position, could be exerting any significant tidal force
  on NGC 1156.

Although there have been many studies on NGC 1156 (especially the \hii regions 
  and CO content; see table 1 below), the distance estimate to this galaxy 
  is only based on two studies \citet{karachentsev96} and \citet{tully88}.
\citet{karachentsev96} estimated the distance to NGC 1156 
  to be $d =7.8 \pm 0.5$ Mpc ($(m-M)_0 =29.46\pm 0.15$ mag)
  using the brightest stars method with $V$ (300 s exposure time) and
  $I$ (300 s) CCD images for the central $80\arcsec \times 120\arcsec$ area
  obtained at the Special Astrophysical Observatory (SAO) 6 m telescope.
From the Tully-Fisher relation,
  \citet{tully88} obtained the distance to NGC 1156 of 6.4 Mpc
  ($(m-M)_0 = 29.02 \pm 0.40$ mag).
There is no estimate for the total (foreground $+$ internal)  
  reddening toward NGC 1156,
  and there are only two foreground reddening estimates to NGC 1156:
  $E(B-V)=0.16$ mag and 0.24 mag measured by \citet{burstein84} and
  \citet{schlegel89}, respectively.

Thanks to the unprecedented resolving power of the {\it Hubble Space Telescope (HST)},
  we are able to resolve individual stars at the center of galaxy NGC 1156.
This allows us to investigate the color-color diagram of the central stars
  of NGC 1156
  and hence make an accurate estimation of the reddening and then
  the distance modulus to this galaxy.
In this paper, we present a new estimate of the distance and (total) reddening
  to NGC 1156 using the photometry from the {\it HST}
  archive imaging data.
This paper is arranged as follows. 
In section 2, we describe the data set used in this study, and 
  we present data reduction and
  transformation into the standard $UBVI$ photometric system
  in section 3.
In section 4, we show the color-magnitude diagrams, and 
  derive the reddening and distance to NGC 1156.
We discuss the results in section 5, and 
  finally summarize the results in section 6.

\section{DATA}
For this study, we used the imaging data sets of $HST$
  Advanced Camera for Surveys (ACS) High Resolution Channel (HRC) on NGC 1156
  available in the $HST$ archive.
While Wide Field Channel (WFC) is the widely used camera of the ACS,
  HRC is one of the three electronic cameras of ACS
  together with Solar Blind Channel (SBC),
  and is designed for high angular resolution imaging and coronagraphy
  for the wavelength range of 2,000 -- 11,000\AA.
The field-of-view of HRC is $26\arcsec \times 29\arcsec$
  ($1024 \times 1024$ SITe CCD), with 
  pixel size of $0.025\arcsec \times 0.028\arcsec$ ($21 \micron$/pixel).

The data used in this study were obtained on 2005 September $5-6$ (UT)
  with F330W, F435W, F550M, F814W, and F658N filters through 
  $HST$ observing program 10609 (P.I.: William Vacca). 
The accumulated exposure times were 1780 s ($4 \times 445$ s) for the F330W band,
  592 s ($4 \times 148$ s) for the F435W band,
  780 s ($4 \times 195$ s) for the F550M band, 
  364 s ($4 \times 91$ s) for the F814W band, and
  240 s for the F658N band, and for this study
  we used the images obtained in the F330W, F435W, F550M, and F814W bands.
Figure 1 displays the grey-scale image of NGC 1156 
  ($5\arcmin \times 5\arcmin$) taken from the Digitized Sky Survey (DSS),
  and figure 2 shows the $HST$ ACS/HRC F814W image of NGC 1156 
  ($26\arcsec \times 29\arcsec$) which is the very central region (inner box) 
  of NGC 1156 in figure 1.

\begin{figure}
\center
\includegraphics[scale=.4]{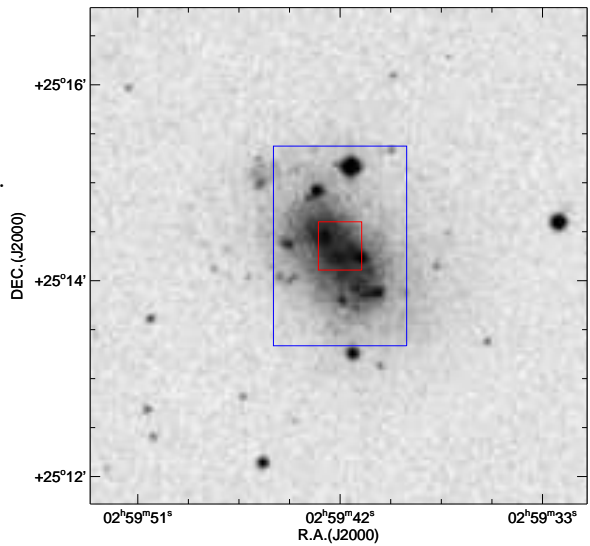}
\caption[1]{Gray-scale second generation
Digitized Sky Survey (DSS) image of NGC 1156.
The size of the field of view is $5\arcmin \times 5\arcmin$.
North is at the top and east is to the left.
The coordinate of NGC 1156 given by the NASA/IPAC Extragalactic Database (NED)
  is ($\alpha_{J2000}$, $\delta_{J2000}$)
  $=$ (02$^h$ 59$^m$ 42.19$^s$, $+25\arcdeg$ $14\arcmin$ 14.2\arcsec),
  which is also the field center in this image.
The inner box denotes the area ($26\arcsec \times 29\arcsec$)
  covered by the $HST$ images in this study (also shown in figure 2),
  and the outer box shows the area ($80\arcsec \times 120\arcsec$)
  covered by  \citet{karachentsev96}.
}
\label{fig1}
\end{figure}

\begin{figure}
\center
\includegraphics[scale=.4]{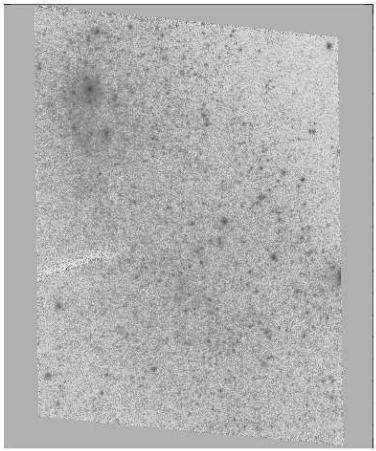}
\caption[2]{Geometric distortion corrected
  $HST$ ACS/HRC F814W image of NGC 1156.
The field of view is about $26\arcsec \times 29\arcsec$,
  which shows the very central region of NGC 1156 shown in figure 1.
North is at the top and east is to the left.
The coordinate of the center is ($\alpha_{J2000}$, $\delta_{J2000}$)
 $=$ (02$^h$ 59$^m$ 42.2$^s$, $+25\arcdeg$ $14\arcmin$ 20.2\arcsec).
The bar feature on the (lower) left is the ``coronagraphic finger''
  of the ACS/HRC.
}
\label{fig2}
\end{figure}

\section{DATA REDUCTION}
\subsection{Photometry}
Photometry of NGC 1156 was carried out for the data taken by $HST$ ACS/HRC
  using the DOLPHOT package developed by \citet{dolphin00a, dolphin00b}.
The DOLPHOT routine performs point spread function fitting to the stars and
  gives the magnitudes in the standard Johnson system
  as well as instrumental magnitudes in the $HST$ filter system.
DOLPHOT uses zeropoints and transformation coefficients of \citet{sirianni05},
  which was recently revised (see $HST$ ACS Web page, 
  http://www.stsci.edu/hst/acs/analysis/zeropoints/).

Figure 3 shows the photometric errors
  from the DOLPHOT package as a function of magnitude.
The DOLPHOT gives five classifications for objects 
  (object 1 : good star; 
  object 2 : possible unresolved binary, two stars combined during photometry iterations; 
  object 3 : bad star, centered on saturated pixel or bad column;
  object 4 : single-pixel cosmic ray or hot pixel;
  object 5 : extended object), and
  here we plot only objects with reliable measurement
  (i.e., flagged as object type 1; $N=5580$).

\begin{figure}
\center
\includegraphics[scale=.4]{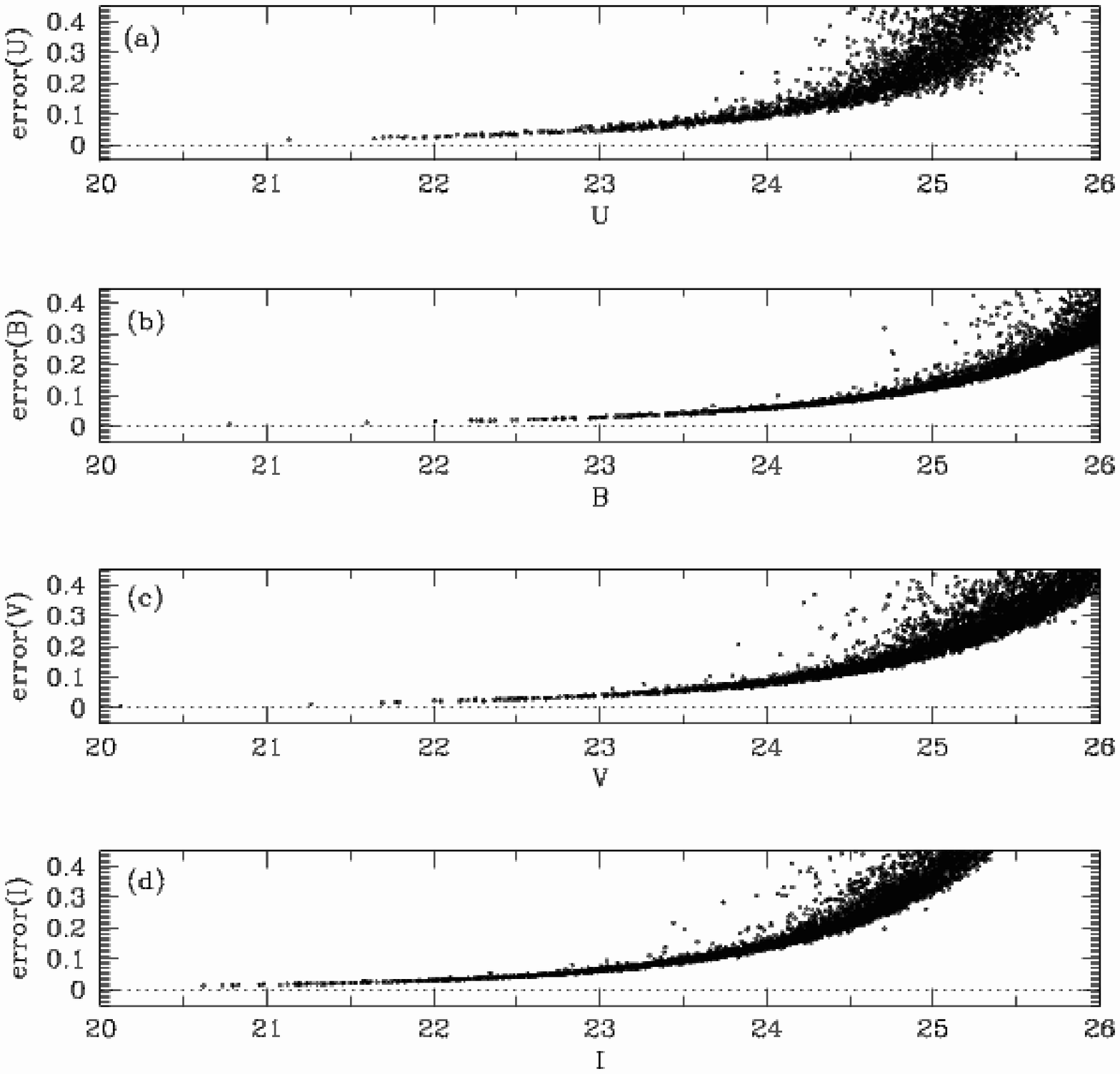}
\caption[3]{Photometric errors from the DOLPHOT package as a function of magnitude.
Only stars with DOLPHOT object type 1 (good star; $N=5580$) are plotted.
} 
\label{fig3}
\end{figure}

\subsection{Transformation to the $UBVI$ system}

We use the $UBVI$ filter system
  for the analysis of the photometric properties and
  determination of the distance to NGC 1156.
Although the DOLPHOT package effectively converts the ACS filter system 
  into the $UBVI$ system
  using the coefficients of \citet{sirianni05}, 
  the transformation from F550M to $V$-band is not provided.
While the recent transformation relations 
  from the $HST$ ACS/WFC system (including F550M filter) 
  to the $BVRI$ photometry 
  were given by \citet{saha11},
  their relations are solely based on the WFC, excluding the HRC.
We, therefore, transform the F550M magnitude into the $V$ magnitude 
  as described below,
  while the conversions from the F330W, F435W, and F814W filters
  to the $U, B$, and $I$ bands, respectively,
  are made by the DOLPHOT package.

First, we transform the F550M magnitudes into the F555W magnitudes
  using the equation 
\begin{equation}
\begin{array}{rcll}
 F555W & = & F550M + 0.232(F550M-F814W)  & \\
       & & ~~~~~~~~~~{\rm for}~ (F550M-F814W) < 1.337~ {\rm mag} \\
 F555W & = & F550M + 0.311               & \\
       & & ~~~~~~~~~~{\rm for}~ (F550M-F814W) \geq 1.337~ {\rm mag}
\end{array}
\end{equation}
  to use the transformation coefficients from F555W to $V$ given by 
  \citet{sirianni05} in the DOLPHOT package.

Equation (1) is derived using the photometry of the stars
  in the globular cluster 47 Tucanae (NGC 104) obtained from the $HST$ ACS/HRC images
  and using the Padova isochrones \citep{marigo08}.
Figure 4 (a) shows the color-color diagram of the stars
  in the globular cluster 47 Tuc obtained from the $HST$ ACS/HRC images
  with F550M, F555W and F814W filters.
Small dots show the stars of 47 Tuc from the DOLPHOT package
  with object type 1 and F550M error $<0.05$ mag.
Among these stars we selected stars with F550M error $<0.006$ mag
  and performed the ordinary least-squares fitting \citep{isobe90}
  with 3-sigma clipping for these stars, which are shown in circles 
  and a blue slanted line denoting the upper part of the equation (1).
Since there are not enough number of stars to use
  at the red part (F550M$-$F814W $\geq 1.337$ mag),
  we have used the Padova isochrones \citep{marigo08} as in figure 4 (b).
\citet{saviane08} find $12 +$ log (O/H) $=8.23$ for NGC 1156 
  and using the equations [Fe/H] $=$ log (O/H) $+3.34 = -0.43$ dex
  \citep{bono10} and log Z $= 0.977$ [Fe/H] $-1.699$ \citep{bertelli94} 
  we assume Z $=0.008$ for NGC 1156.
The various ages (1, 2, 5, and 12 Gyr) used in figure 4 (b)
  give little differences and
  we have used a constant value of 0.311 (lower part of the equation (1)) 
  for the red part of (F550M$-$F814W) $\geq 1.337$ mag.
Figure 4 (c) is the composite of the panels (a) and (b), showing
  the stars in the globular cluster 47 Tuc (dots for the stars with 
  F550M error $<0.05$ mag and small circles for those with 
  F550M error $<0.006$ mag),
  the Padova isochrones with Z $=0.008$
  and ages of 1 Gyr (triangles) and 12 Gyr (large circles), and
  the equation (1) plotted in solid line.

\begin{figure*}
\center
\includegraphics[scale=.7]{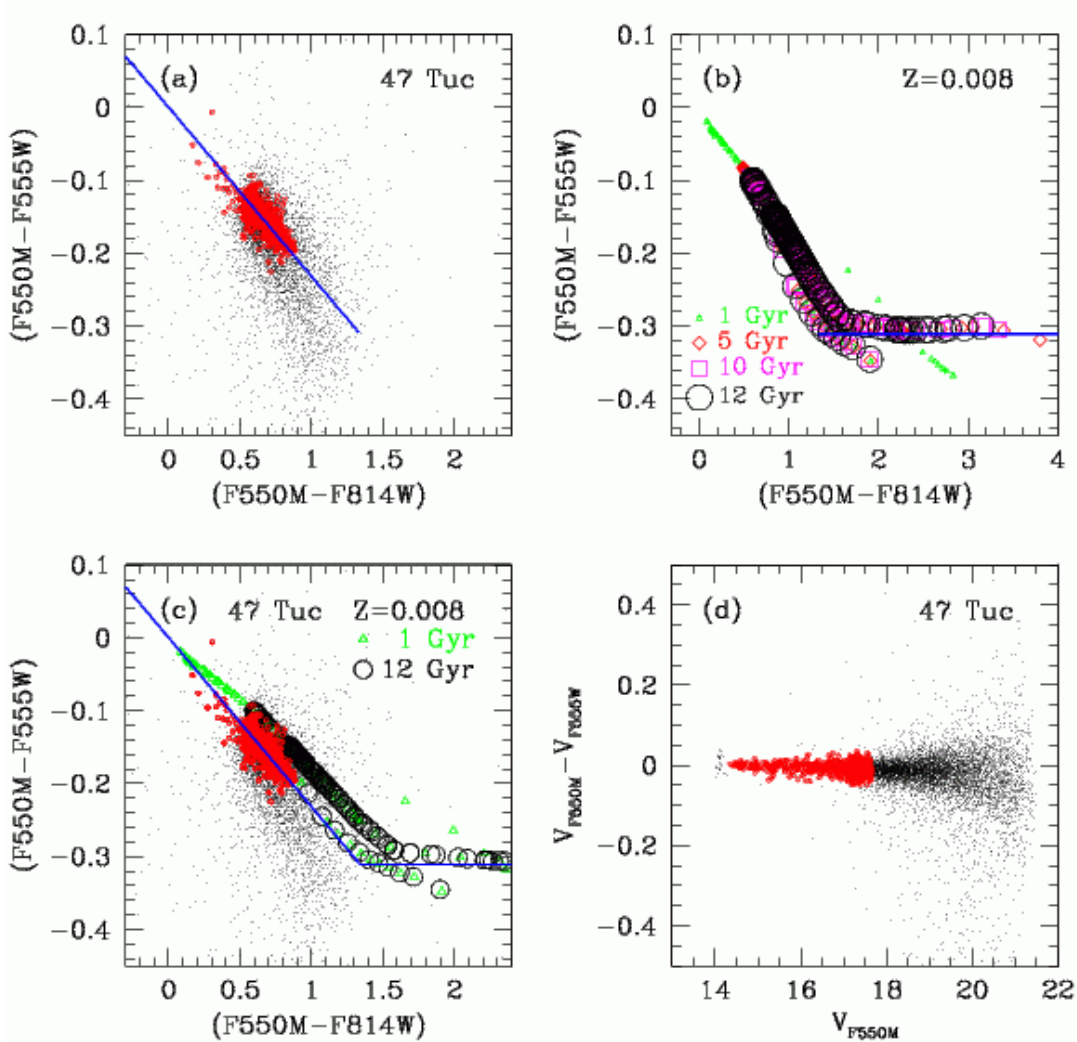}
\caption[4]{(a) The color-color diagram of the stars
  in the globular cluster 47 Tuc (NGC 104) obtained
  from the $HST$ ACS/HRC images.
Small dots are for the stars with F550M error $<0.05$ mag.
Stars with F550M error $<0.006$ mag are selected and used for
  the ordinary least-squares fitting \citep{isobe90}
  with 3-sigma clipping, which are shown in circles 
  and slanted line 
  for the blue part of (F550M$-$F814W) $\simlt 1.3$.
(b) The Padova isochrones \citep{marigo08}
  with Z$=0.008$ and various ages (1, 5, 10, and 12 Gyr) are shown.
The constant value determined at (F550M$-$F814W) $\simgt 1.3$ 
  is shown as a horizontal line.
(c) Composite of panels (a) and (b),
  showing the stars of the globular cluster 47 Tuc
  (dots for the stars with F550M error $<0.05$ mag and
  small circles for the stars with F550M error $<0.006$ mag)
  and the Padova isochrones with Z$=0.008$ and ages of 1 Gyr (triangles)
  and 12 Gyr (large circles).
The solid line shows the selected transformation
  with slope of 0.232 (for the blue part of F550M$-$F814W $\leq 1.337$ mag)
  and 0 (for the red part of F550M$-$F814W $\geq 1.337$ mag).
(d) The difference between the two $V$ magnitudes transformed from F550M and
  from F555W, which shows very good agreement especially for the stars
  with small errors (F550M error $<0.006$ mag) shown in circles. 
}
\label{fig4}
\end{figure*}

To test the efficiency of this method, we compare the $V$ magnitudes
  transformed from F550M and from F555W
  for the stars in the globular cluster 47 Tuc in figure 4 (d),
  where circles are for the stars with F550M error $<0.006$ mag
  shown also as circles in panel (a). 
The two $V$ magnitudes transformed from F550M and from F555W
  are in very good agreement with each other 
  as seen in figure 4 (d).

\section{RESULTS}

\subsection{Color-Magnitude Diagrams}

Figure 5 displays the color-magnitude diagrams (CMDs) for the measured stars
  (N$= 5248$) located in the central region of NGC 1156,
  only with DOLPHOT object type of 1.
While blue supergiant (BSG)/blue main sequence stars are clearly seen
  in the $(V, U-B)$ (figure 5(a)) and $(V, B-V)$ (figure 5(b)) CMDs
  reaching $B=21.6$ mag and $V=21.3$ mag,
  the ($(V, V-I)$ (figure 5(c)) and $(I, V-I)$ (figure 5(d)) CMDs
  show both the BSG and red supergiant (RSG) stars well,
  the color of the latter being $(V-I) \sim 2$ mag
  and reaching up to $V=22.7$ mag and $I=20.6$ mag.
Red giant branch (RGB) stars are located below the tip of the RGB (TRGB),
  which might be much fainter than $I \sim 24$ (see figure 5(d)).
There should be many asymptotic giant branch (AGB) stars between
  the RSG and RGB stars (see, e.g., \citet{cioni05}).

\begin{figure*}
\center
\includegraphics[scale=.8]{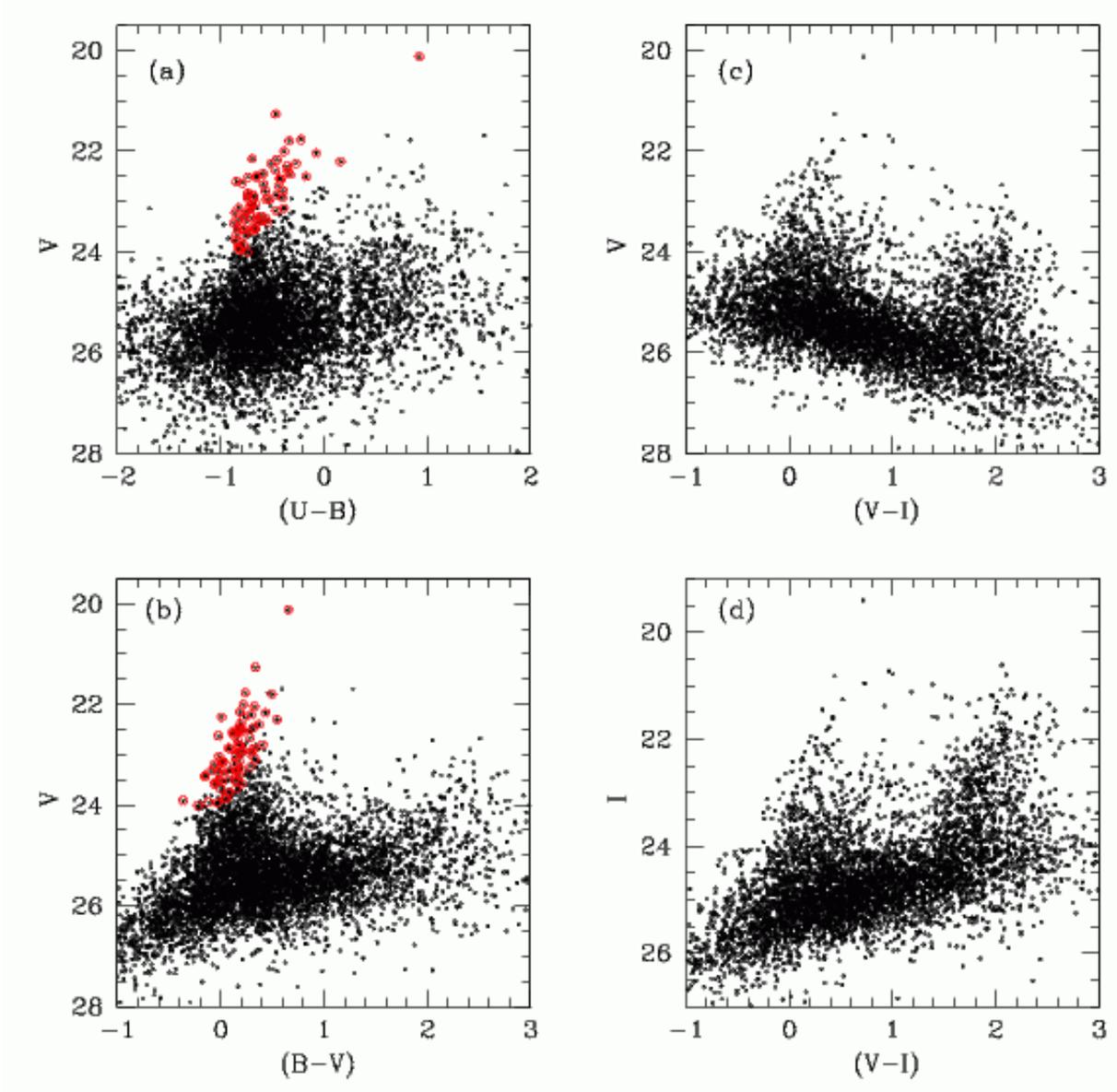}
\caption[5]{Color-magnitude diagrams (CMDs) of NGC 1156
  obtained from the $HST$ ACS/HRC images.
Only the objects of type 1 (good star) are plotted
  from the DOLPHOT photometry.
The large circles in panels (a) and (b) are the objects  
  with $U$ photometric errors smaller than 0.05 mag
  which are used to estimate the total reddening value toward NGC 1156
  in figure 6 below.
}
\label{fig5}
\end{figure*}

If we assume (i) the distance to NGC 1156 to be $d = 7.1 \pm 1.0$ Mpc 
  ($(m-M)_0 = 29.24 \pm 0.31$ mag) given by the NASA/IPAC Extragalactic Database (NED)
  which is the mean of the two values from \citet{tully88} ($(m-M)_0 = 29.02 \pm 0.40$ mag) 
  and \citet{karachentsev96} ($(m-M)_0 = 29.46 \pm 0.15$ mag), 
  (ii) the total reddening toward NGC 1156 to be $E(B-V) =0.35 \pm 0.05$ mag 
  as obtained in the following subsection, and
  (iii) the $I$-band absolute magnitude of the TRGB to be 
  $M_{I, TRGB} \approx -4.0 \pm 0.1$ mag \citep{lee93},
  then the $I$-band magnitude of the TRGB might be located at $I \sim 25.8$ mag.
The $U$-magnitude calibration from \citet{sirianni05} is bifid 
  above and below $(U-B) = 0.2$ mag (see their table 23). 
At the very color of $(U-B) = 0.2$ mag, the derived $U$-magnitudes are
  different by 0.1 mag (see also their figure 22),
  and this appears to be the reason for the vertical gap-like feature 
  at $(U-B) = 0.2$ mag in the $(V, U-B)$ CMD (figure 5(a)).
The cross identification results of the bright stars and star clusters
  in the $HST$ images and the Canada-France-Hawaii Telescope near-infrared
  images will be shown in a subsequent paper (J. Kyeong et al. in preparation).

\subsection{Interstellar Reddening toward NGC 1156}
Since NGC 1156 is a dIrr galaxy containing many star forming regions \citep{barazza01},
  the internal reddening in NGC 1156 might not be negligible
  and the estimate of the total\footnote{`foreground' (found in 
  the Milky Way) + `internal' (found in the program galaxy)}
  reddening toward NGC 1156 is important for the photometric studies of this galaxy.
However, there are only two foreground reddening values toward NGC 1156.
\citet{burstein84} measured $A_B=0.66$ mag using 21 cm \hi column
  density and faint galaxy counts method. 
On the other hand, \citet{schlegel89}
  determined $A_B=0.968$ mag based on the $COBE$/DIRBE measurement 
  of diffuse infrared emission. 
By adopting $R_V=3.1$ and the interstellar extinction law
  of \citet{cardelli89},
  these values convert to $E(B-V)=0.16$ mag and 0.24 mag, respectively.

Using the color-color diagram produces a robust method
  for determining the interstellar reddening value
  only if $UBV$ photometric data are available.
Fortunately, our four filter photometry allows us to derive 
  the interstellar reddening value using the $(U-B, B-V)$ color-color diagram,
  which is shown in figure 6.
The total reddening value,
  that is foreground reddening plus internal reddening inside the galaxy,
  is derived shifting the zero age main sequence (ZAMS) relation
  given by the Padova isochrone (log $t=6.0$, Z$=0.008$;
  \citet{marigo08}, \citet{bertelli94}, \citet{saviane08}) 
  along the reddening line of $E(U-B) =0.72 E(B-V)$ \citep{gunn83}.
Only the stars with $U$ photometric errors smaller than 0.05 mag and
  object type 1 are included for the fit,
  which are mostly in the blue main sequence region in the CMD
  as shown in the panels (a) and (b) of figure 5 as large (red) circles.
The photometric errors for these stars in $B$ and $V$ bands are
  less than $0.06$ mag and $0.09$ mag, respectively.

\begin{figure*}
\center
\includegraphics[scale=.7]{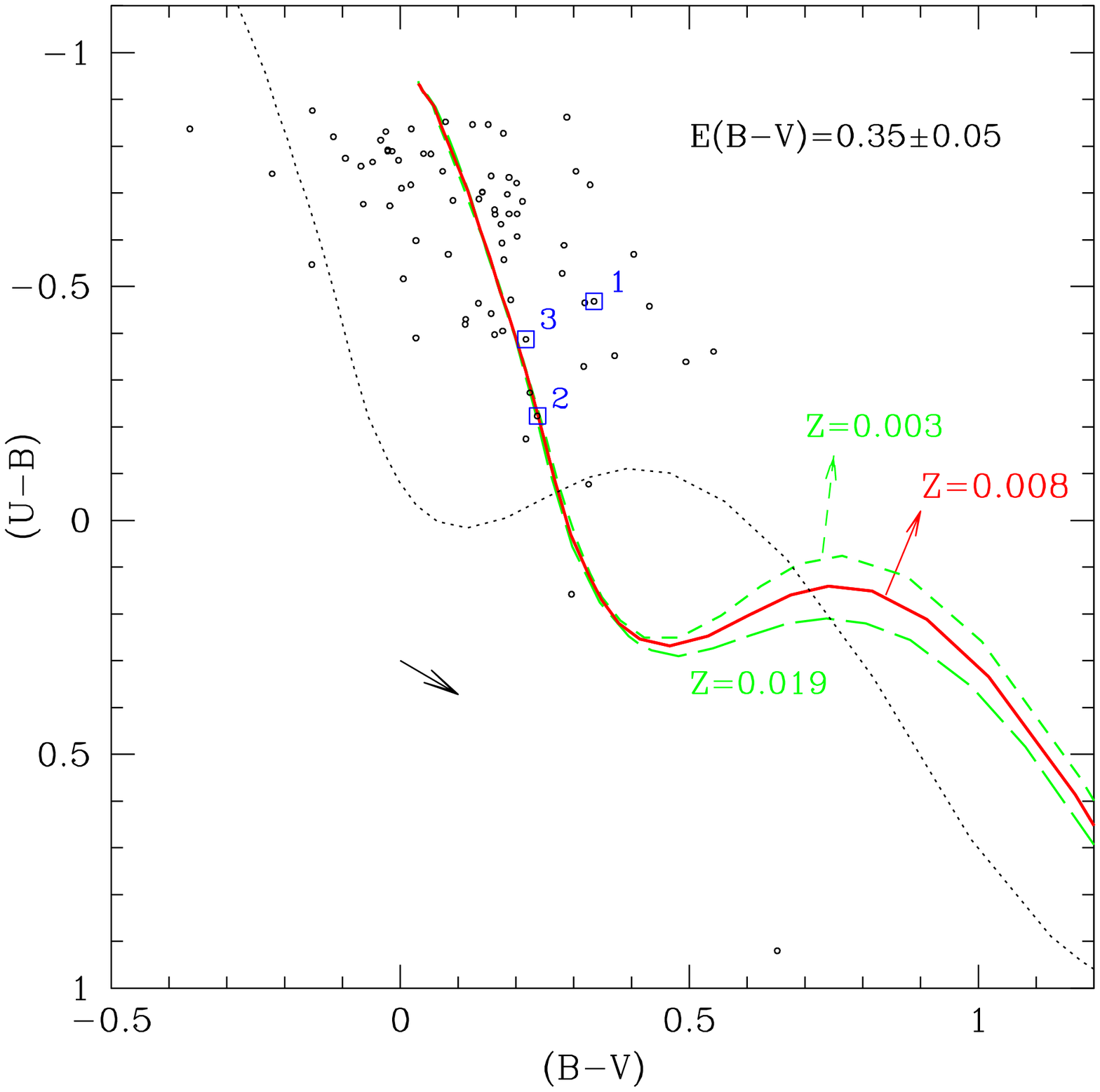}
\caption[6]{($U-B, B-V$) color-color diagram of the stars in
  NGC 1156.
The objects with $U$ photometric errors smaller than 0.05 mag and object type 1
  are plotted as open circles.
The thin dotted line represents the unreddened ZAMS line given by
  the Padova isochrone \citep{marigo08} with log $t = 6.0$, Z$=0.008$
  \citep{saviane08}, and
  the thick solid line represents the ZAMS line reddened by the amount
  of $E(B-V)=0.35 \pm 0.05$ mag.
The ZAMS lines for Z$=0.003$ and Z$=0.019$ (both for log $t = 6.0$)
  reddened by $E(B-V)=0.35$ mag
  are also shown in thick short-dashed and long-dashed lines, respectively, for comparison,
  where the chosen metallicity gives little difference in obtaining
  the reddening value.
The arrow at $(B-V) \sim 0$ mag is the reddening vector with $E(B-V)=0.10$ mag.
The three, numbered stars with boxes are the BSGs selected in section 4.3 
  and shown in table 2 in the order named.
}
\label{fig6}
\end{figure*}

The best fit in figure 6 yields $E(B-V) =0.35 \pm 0.05$ mag for the total reddening value
  of NGC 1156, which will be used
  in the derivation of the distance in the next subsection. 
Since there are several very young star clusters (log $t < 6.6$) in the core region
  of NGC 1156, the internal reddening that contributed to this value might be
  due to the gas and dust associated with the star forming regions.
The extinction values are calculated for a total-to-selective extinction
  ratio of $A_V /E(B-V)=3.1$ using the equations given by 
  \citet{cardelli89}:
  $A_B = 4.14 E(B-V) = 1.45$ mag, 
  $A_V =  3.1 E(B-V) = 1.09$ mag, and
  $A_I = 1.48 E(B-V) = 0.52$ mag.

\subsection{Distance to NGC 1156}
\citet{bottinelli84} determined the distance 
  to NGC 1156 as 3.9 Mpc ($(m-M)_0 = 27.98$ mag) by the $B$-band Tully-Fisher relation
  with a relatively large error ($\sim 1.2$ mag) in the distance modulus 
  \citep{karachentsev96}.
\citet{tully88} obtained a larger distance of 6.4 Mpc ($(m-M)_0 = 29.02 \pm 0.40$ mag) 
  again from the Tully-Fisher relation and a heliocentric velocity of 372 \kms.
Using the magnitudes of RSG and BSG candidates,
  \citet{karachentsev96} obtained the distance of 
  $d =7.8 \pm 0.5$ Mpc ($(m-M)_0 =29.46\pm 0.15$ mag) to NGC 1156,
  which is the mean value of $(m-M)_0 =29.61$ mag (obtained from the three
  RSGs with a mean $V$-band magnitude of $\langle V(3R) \rangle = 22.45$ mag) and
  $(m-M)_0 =29.32$ mag (obtained from the three BSGs with 
  $\langle B(3B) \rangle = 21.31$ mag).
They searched for RSG and BSG candidates 
  only outside of the central crowded areas to avoid the crowding in the inner regions
  and obtained a mean apparent magnitude for the three brightest red stars of
  $\langle V(3R) \rangle = 22.45$ mag, and that for the three brightest blue stars
  of $\langle V(3B) \rangle = 21.24$ mag.

With non-variable absolute visual magnitudes up to $M_V \simeq -9.5$ mag,
  the brightest stars method is quite useful in estimating the distances to galaxies
  without the necessity of repeated observations \citep{hubble36, humphreys87,
  sandage88, karachentsev94, rozanski94, lyo97, kudritzki03, bresolin03, vaduvescu05,
  kudritzki08, kudritzki10}.
Instead of the single brightest star, the average magnitude of the three
  brightest stars have been used to minimize the effects of misidentifying
  the brightest individual star and 
  to reduce the stochastic effect in obtaining the mean luminosity
  of the brightest stars \citep{rozanski94, lyo97}.
While \citet{rozanski94} claimed rather larger errors
  in this method of 0.58 mag and 0.90 mag, respectively, for the brightest
  red and blue stars,
  \citet{karachentsev94} suggested much smaller errors of
  0.30 mag and 0.45 mag for the brightest red and blue stars, respectively.
Using new CCD-based data for 17 galaxies, \citet{lyo97} showed that 
  the uncertainties in the distance moduli determined by 
  the brightest red and blue stars are 0.37 mag and 0.55 mag, respectively,
  concluding that the brightest RSGs might be useful in 
  determining the distances to resolved late-type galaxies.
They proposed new calibration equations :
\begin{equation}
\langle M_V(3)_{RSG} \rangle = 0.21 M_B^T - 3.84, ~~\sigma(M_V)=0.37 ~{\rm mag}
\end{equation}
and
\begin{equation}
\langle M_B(3)_{BSG} \rangle = 0.30 M_B^T - 3.02, ~~\sigma(M_B)=0.55 ~{\rm mag}.
\end{equation}

In order to make it free from
  any contamination by foreground stars and
  to select the three brightest blue and red stars, we consider stars
  bluer than $B-V = 0.4$ mag (in $(B, B-V)$ CMD) and redder than $B-V = 2.0$ mag
  (in $(V, B-V)$ CMD), respectively \citep{rozanski94}.
In figure 7, we show the selected supergiant stars:
  pentagons are the three brightest blue stars and
  the open circles are the three brightest red stars,
  while the triangles and squares in panels (c) and (d) are the brightest blue and
  red, respectively, stars selected in \citet{karachentsev96}.
Table 2 lists the equatorial coordinates and photometry results of 
  the selected BSG and RSG stars.
While we denote $V(3)$ as the average $V$-band magnitude
  of the three brightest stars selected in the $V$-band,
  it is worth noting here that 
  the brightest RSG stars in $V(3)$ or $K(3)$ will not, in general, 
  be the same stars as in $I(3)$ or $R(3)$, 
  and this situation is the same
  for the brightest BSGs \citep{rozanski94}.
It is natural that similar but not the same stars are selected
  between our study and that of \citet{karachentsev96}
  considering the different areas of the studies:
  we use the $HST$ images of the central 
  $26\arcsec \times 29\arcsec$ region of NGC 1156,
  while \citet{karachentsev96} searched for supergiant candidates
  only outside the central/crowded areas in their $80\arcsec \times 120\arcsec$ 
  CCD images.
The mean magnitudes of the three brightest BSGs are
  $\langle B(3B) \rangle = 21.943$ mag and $\langle V(3B) \rangle = 21.680$ mag
  and the mean color is $\langle (B-V)(3B) \rangle = 0.263$ mag,
  where 3B denotes the three brightest BSGs.
The redder $(V-I)$ colors of the BSGs in this study ($\langle (V-I) \rangle = 0.392$ mag)
  over those of the BSGs in \citet{karachentsev96} 
  ($\langle (V-I) \rangle = 0.077$ mag) might be due to the crowding of
  more dust and star forming regions in the central area of NGC 1156
  that caused more reddening.
The mean magnitudes of the three brightest RSGs in this study
  ($\langle V(3R) \rangle = 22.758$ mag, $\langle I(3R) \rangle = 20.871$ mag) are somewhat
  fainter than those of the RSGs in \citet{karachentsev96}
  ($\langle V \rangle = 22.443$ mag, $\langle I \rangle = 20.317$ mag) :
  $\langle V \rangle_{this ~study} - \langle V \rangle_{Karachentsev} = 0.315$ mag and
  $\langle I \rangle_{this ~study} - \langle I \rangle_{Karachentsev} = 0.554$ mag.

\begin{figure*}
\center
\includegraphics[scale=.7]{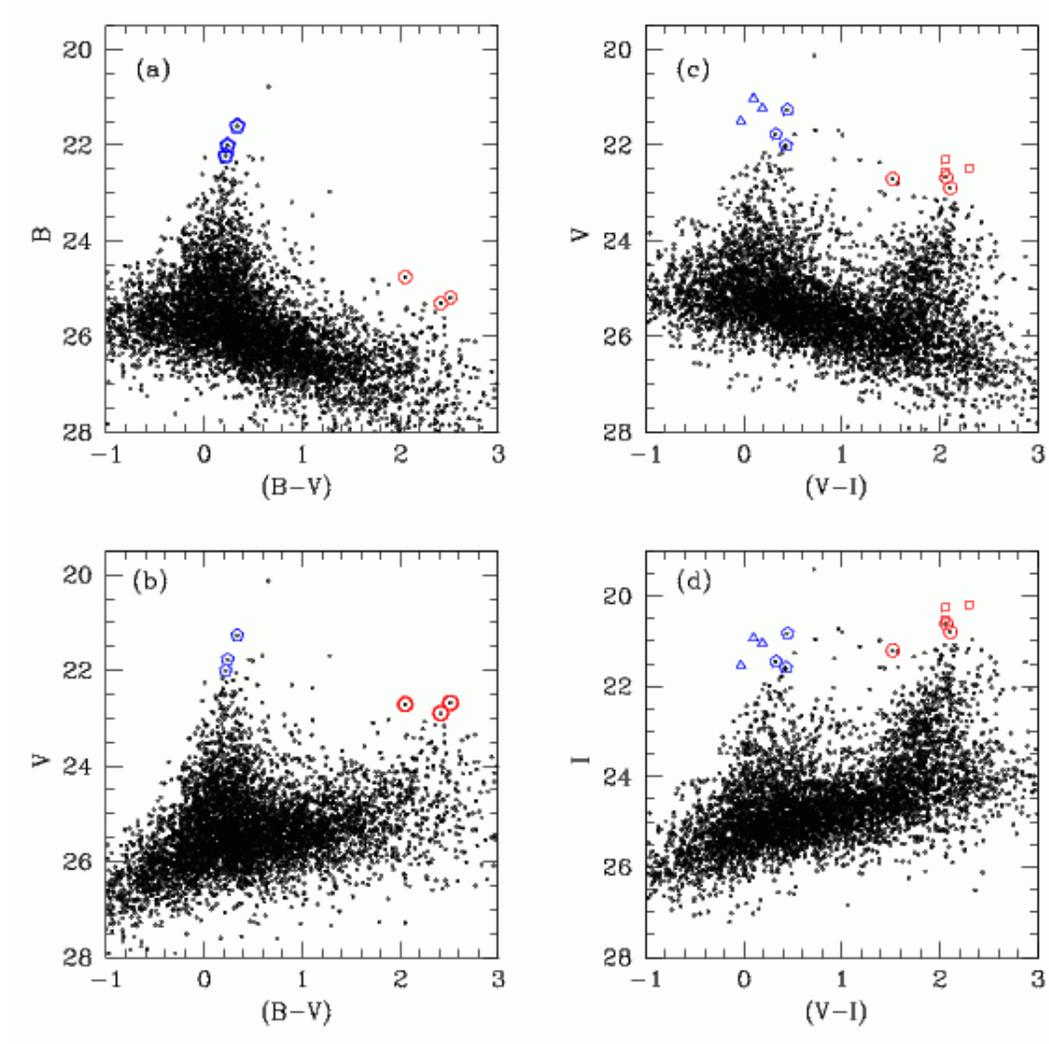}
\caption[7]{The location of the selected blue supergiants (BSGs)
  and red supergiants (RSGs) in (a) $(B, B-V)$ CMD, (b) $(V, B-V)$ CMD,
  (c) $(V, V-I)$ CMD, and (d) $(I, V-I)$ CMD.
Pentagons and open circles are the BSG and RSG stars, respectively,
  selected in this study, while
  triangles and squares in panels (c) and (d) are the brightest blue and
  red, respectively, stars selected in \citet{karachentsev96}.
BSGs selected in the $(B, B-V)$ CMD (panel (a)) and
  RSGs selected in the $(V, B-V)$ CMD (panel (b)) are denoted boldly
  in each panel.
}
\label{fig7}
\end{figure*}

Using the total blue absolute magnitude of NGC 1156 ($M_B^T = -18.64$ mag)
  given by \citet{barazza01} (we reddened $M_{B,0}^T$ into $M_B^T$ again
  using $A_B = 0.96$ mag as given in their study)
  and equation (2), we obtain the mean total visual absolute magnitude of
  $\langle M_V(3)_{RSG} \rangle = -7.75$ mag.
Using the average $V$ magnitude of the three brightest RSGs 
  derived in the previous paragraph ($\langle V(3R) \rangle = 22.758$ mag), 
  we get the distance modulus of $(m-M)_0 = 22.758 + 7.75  - 3.1 E(B-V) = 29.423$ mag.
Using this value, iteratively, to derive the total blue absolute magnitude
  of NGC 1156 assuming $B_T = 11.78$ mag \citep{barazza01} and
  the total reddening of $A_B = 1.45$ mag (section 4.2),
  we get $M_B^T = -19.093$ mag.
Inserting this value into the equation (2) and iterating this process
  until the values converge, we obtain
  $M_B^T = -19.22$ mag, $\langle M_V(3)_{RSG} \rangle = -7.88$ mag, and
  the distance modulus $(m-M)_0 = 29.55$ mag ($d = 8.1$ Mpc).

Similarly, using the total blue absolute magnitude of NGC 1156 ($M_B^T = -18.64$ mag)
  given by \citet{barazza01} and equation (3),
  we get the mean total blue absolute magnitude of
  $\langle M_B(3)_{BSG} \rangle = -8.61$ mag.
Using the average $B$ magnitude of the three brightest BSGs
  derived above ($\langle B(3B) \rangle = 21.943$ mag),
  we get the distance modulus of $(m-M)_0 = 21.943+8.61 -4.14 E(B-V) =29.104$ mag.
Using this value to derive again the total blue absolute magnitude
  of NGC 1156 and iterating the process until the values converge,
  we get $M_B^T = -18.83$ mag, $\langle M_B(3)_{BSG} \rangle = -8.67$ mag, and
  the distance modulus $(m-M)_0 = 29.16$ mag ($d = 6.8$ Mpc).

The average of the distance moduli obtained from using the BSGs and RSGs
  is $(m-M)_0 = 29.36 \pm 0.20$ mag ($d = 7.4 \pm 0.7$ Mpc).
Since the uncertainties of the distance estimates from these two methods
  are different, we use different weights for each of them.
The errors in the distance estimation methods using BSGs and RSGs, respectively, are
  0.90 mag : 0.58 mag (1.55 : 1) in the study of \citet{rozanski94},
  0.45 mag : 0.30 mag (1.50 : 1) in that of \citet{karachentsev94}, and
  0.55 mag : 0.37 mag (1.49 : 1) in that of \citet{lyo97}.
The mean values of the three estimates of the errors for the BSGs and RSGs are
  0.63 mag and 0.42 mag, respectively, and the ratio is 1.50 : 1.
Giving single weight to the distance estimate from the BSGs and
  a weight of 1.5 to that from the RSGs, we finally obtain
  $(m-M)_0 = 29.39 \pm 0.20$ mag ($d = 7.6 \pm 0.7$ Mpc).
This result is in very good agreement with that derived by \citet{karachentsev96},
  while somewhat larger than that of \citet{tully88}.
The fact that the distance estimate of this study is very similar to that
  of \citet{karachentsev96} could be considered reasonable 
  because the magnitudes of the selected BSGs and RSGs are not much different and
  the studied areas of the two studies are located close to each other.

\section{DISCUSSION}
While \citet{devaucouleurs91} lists the angular diameter of NGC 1156
  to be $3.3\arcmin$,
  our study and that of \citet{karachentsev96}
  have used, respectively, only 
  the $HST$ data of the central $26\arcsec \times 29\arcsec$ area and
  the SAO 6 m telescope data of the central $80\arcsec \times 120\arcsec$ area.
This means the present study covered only 2.5\% of the area of NGC 1156 
  and that of \citet{karachentsev96} 31.2\%.
In fact, in order to select the true brightest stars in a galaxy,
  we would need to observe the entire area of the galaxy.
It, therefore, might be easily speculated that there could be 
  brighter stars in NGC 1156 than the brightest stars selected 
  in the two studies above.

Figure 1 shows that the area covered by \citet{karachentsev96}
  includes most of the bar-like main body and most of the star forming regions
  of NGC 1156, and it seems there may not be 
  many, if any, brighter stars outside of the area
  studied by \citet{karachentsev96}.
The fact that our study have used a smaller area than that of 
  \citet{karachentsev96} is consistent with 
  somewhat fainter $\langle V \rangle$ magnitudes for both 
  the BSGs ($\Delta V = V_{This~ study}-V_{Karachentsev} \sim 0.44$ mag) and
  the RSGs ($\Delta V \sim 0.32$ mag) selected in this study
  as shown in figure 7 (c).
The $\langle I \rangle$ magnitude for the RSGs selected in this study
  is quite fainter than that of \citet{karachentsev96}
  ($\Delta I = I_{This~ study}-I_{Karachentsev} \sim 0.55$ mag), but
  the $\langle I \rangle$ magnitude for the BSGs selected in this study
  is only a little fainter than that of \citet{karachentsev96}
  ($\Delta I \sim 0.12$ mag; figure 7 (d)).
Although the smaller area covered by the $HST$ images in this study
  compared to that of \citet{karachentsev96} renders fainter mean magnitudes
  of the brightest stars,
  the new calibration for the brightest stars (from \citet{lyo97})
  and the new estimate of the total reddening to NGC 1156 used in this study 
  give almost the same distance.
It is still possible, though, that 
  we might get brighter magnitudes for the brightest stars (and smaller distance modulus),
  if we observe a larger area than 
  that of \citet{karachentsev96} ($r \simgt 0.9\arcmin$).

The study of \citet{karachentsev96} and our study have used
  different observation data and 
  different regions to search for the brightest stars:
  the former focused only on the outer regions 
    in their $80\arcsec \times 120\arcsec$ images to avoid the central crowded areas
  and the latter used only the central $26\arcsec \times 29\arcsec$ area of NGC 1156.
Therefore, we can obtain more reliable results if we combine the data
  on the brightest stars from these two studies. 
The selection of three brightest RSGs among the six RSGs in the two studies
  give the same stars as in the study of \citet{karachentsev96}
  since all of their stars are brighter than those of the present study.
Using the three RSGs of \citet{karachentsev96} with 
  $\langle V(3R) \rangle = 22.45$ mag and the equation (2), 
  we obtain the distance modulus of $(m-M)_0 = 29.16$ mag ($d = 6.8$ Mpc).

\citet{karachentsev96} used only $V$ and $I$ bands for their study.
They used these two bands to search for both BSGs and RSGs
  and they obtained the mean $B$ magnitude of the three BSGs 
  ($\langle B(3B) \rangle = 21.31$ mag) from conversion 
  of the mean $V$ magnitude and $(V-I)$ color 
  ($\langle V(3B) \rangle = 21.24$ mag, $\langle (V-I) \rangle = 0.08$ mag).
Although the $B$ magnitude of each of the three BSGs in the study of 
  \citet{karachentsev96} is not known,
  we might be able to assume that they are all brighter than 
  any of the BSGs used in the present study
  considering the larger magnitude difference in $V$ (figure 7 (c)) than
  that in $I$ (figure 7 (d)) : $\Delta V = \langle V_{This~ study} \rangle
  -\langle V_{Karachentsev} \rangle \sim 0.44$ mag 
  and $\Delta I = \langle I_{This~ study} \rangle
  -\langle I_{Karachentsev} \rangle \sim 0.12$ mag.
Using the three brightest BSGs of \citet{karachentsev96}
  among the six BSGs in the two studies and the equation (3),
  we get $\langle B(3B) \rangle = 21.31$ mag, which gives us
  the distance modulus of $(m-M)_0 = 28.26$ mag ($d = 4.5$ Mpc).

The weighted mean of the distance estimates from using BSGs and RSGs
  using weights of 1 and 1.5, respectively, is 
  $(m-M)_0 = 28.80 \pm 0.20$ mag ($d = 5.8 \pm 0.5$ Mpc),
  while the unweighted mean 
  is $(m-M)_0 = 28.71 \pm 0.50$ mag ($d = 5.5 \pm 1.4$ Mpc).
Using the revised calibration of the BSGs and RSGs in \citet{lyo97}
  and the newly obtained total reddening to NGC 1156 in this study,
  we derive a distance to this galaxy smaller than those obtained 
  in section 4.3 and in \citet{karachentsev96}.
Future wide-field imaging of NGC 1156 at least in $B$ and $V$-bands
  might be helpful in resolving this issue and 
  determining more accurate distance to NGC 1156.

The mean magnitudes and color of the BSGs obtained in this study are
  $\langle V(3B) \rangle = 21.68$ mag,
  $\langle I(3B) \rangle = 21.29$ mag, and
  $\langle (V-I)(3B) \rangle = 0.39$ mag, and 
those of the RSGs obtained in this study are
  $\langle V(3R) \rangle = 22.76$ mag,
  $\langle I(3R) \rangle = 20.87$ mag, and
  $\langle (V-I)(3R) \rangle = 1.89$ mag.
On the other hand, the mean magnitudes and color of the BSGs obtained 
  by \citet{karachentsev96} are
  $\langle V(3B) \rangle = 21.24$ mag,
  $\langle I(3B) \rangle = 21.17$ mag, and
  $\langle (V-I)(3B) \rangle = 0.08$ mag, and
those of the RSGs obtained by \citet{karachentsev96} are
  $\langle V(3R) \rangle = 22.44$ mag,
  $\langle I(3R) \rangle = 20.32$ mag, and
  $\langle (V-I)(3R) \rangle = 2.13$ mag.
All the magnitudes obtained in our study are $0.12 - 0.55$ mag fainter than those
  of \citet{karachentsev96},
while the situation is different for the $(V-I)$ color :
  the mean $(V-I)$ color for the BSGs obtained in our study 
  is 0.32 mag redder than that of \citet{karachentsev96},
  while that for the RSGs is 0.24 mag bluer than that of \citet{karachentsev96}.
It could be possible that there occurs larger absorption
  in the central part of NGC 1156, which affects more for the BSGs 
  than the RSGs.
Nevertheless, since we accurately estimated the reddening value in the central part
  of the galaxy and applied it in determining the distance,
  the distance values obtained in this study are not affected much
  by the differences of reddening values in our study and in the study of 
  \citet{karachentsev96}.

All the studies performed in the 1990s, including those of
  \citet{karachentsev94}, \citet{rozanski94}, and
  \citet{lyo97}, used the $(B-V)$ color to select both BSGs and RSGs.
The wide use of CCDs since then have produced colors using longer wavelengths
  such as $(V-I)$ thanks to the good sensitivities of CCD detectors
  at these wavebands.
It might be helpful if we could use this $(V-I)$ color in selecting
  or investigating the RSGs rather than the $(B-V)$ color.
Using the $(V-I)$ colors for the RSGs assembled in \citet{lyo97}
  together with those for NGC 1156 selected in this study,
  we plot the $(V, V-I)$ and $(I, V-I)$ CMDs in figure 8
  to show the $(V-I)$ color distribution of the RSGs.
While many RSGs are gathered near $(V-I) \sim 2$ mag,
  the whole color range of the RSGs is $1.5 mag \simlt (V-I) \simlt 3.5$ mag.

\begin{figure*}
\center
\includegraphics[scale=.7]{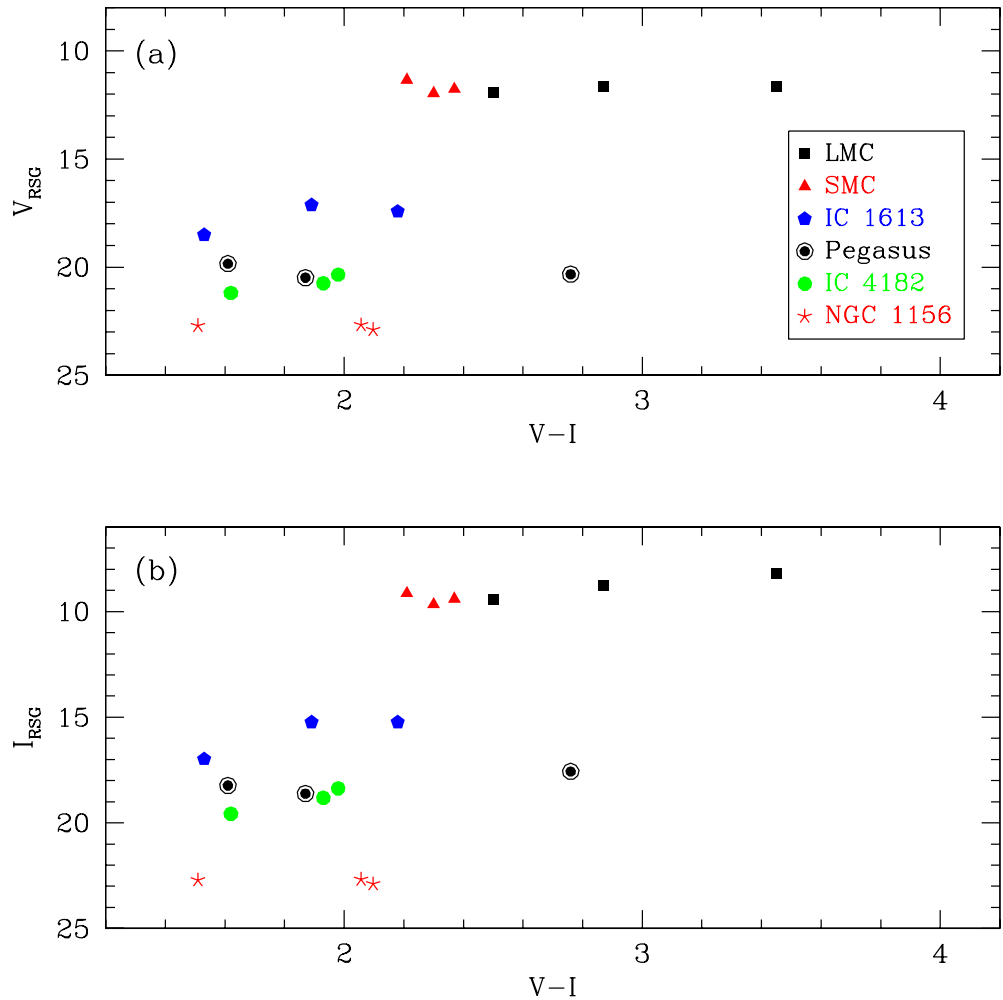}
\caption[8]{The (a) $(V, V-I)$ and (b) $(I, V-I)$ CMDs showing
  the RSGs assembled and used in \citet{lyo97}
  and those selected in this study for NGC 1156.
The symbols for the six galaxies including NGC 1156
  are shown in box in panel (a).
}
\label{fig8}
\end{figure*}

Figure 7 shows that among the three selected RSGs, 
  RSG2 is bluest both in $(B-V)$ and $(V-I)$ colors.
The $(V-I)$ color of this star ($(V-I) = 1.51$ mag) is 
  almost at the blue edge 
  in the color range shown in figure 8.
If we ignore this star and select again three RSGs in 
  figure 7(c), the somewhat fainter and redder star 
  with $V = 22.8$ mag and $(V-I) = 1.56$ mag will be chosen,
  and this does not much affect 
  the resultant distance estimate to NGC 1156.

\section{SUMMARY}

Using the $UBVI$ archive $HST$ ACS/HRC images of the dIrr galaxy NGC 1156,
  we performed DOLPHOT photometry, constructed various combinations of
  CMDs, and estimated the total (foreground $+$ internal) reddening
  and distance to this galaxy.
Although the CMDs are not deep enough to detect the TRGB
  due to the short exposure times of the $HST$ images,
  they are good enough to draw a $(U-B, B-V)$ color-color diagram
  to determine the total interstellar reddening of $E(B-V) =0.35 \pm 0.05$ mag
  and to select the three BSGs and three RSGs,
  which allowed us to determine the distance to this galaxy, 
  $d = 7.6 \pm 0.7$ Mpc ($(m-M)_0 = 29.39 \pm 0.20$ mag).

The CMDs obtained in this study are quite similar to those of
  NGC 6822 \citep{gallart96} (figure 15, $(V, B-V)$ CMD and
    figure 17, $(I, V-I)$ CMD),
  NGC 3109 \citep{minniti99} (figure 2, $(I, V-I)$ and $(V, V-I)$ CMDs),
  WLM \citep{minniti97} (figure 5, $(V, V-I)$ and $(I, V-I)$ CMDs), and/or
  IC 1613 \citep{freedman88} (figure 4, $(V, B-V)$ CMD and
    figure 5, $(I, V-I)$ CMD).
Since NGC 1156 is located much farther than these late-type dwarf galaxies,
  it is necessary to get deeper imaging of NGC 1156
  to obtain deep enough photometry 
  so that we can estimate the distance more reliably using the TRGB method or 
  perform a detailed study of the stellar populations of this galaxy.

\vskip 10mm
The authors are thankful to the anonymous referee for useful comments
  that improved the manuscript.
H.S.P. was supported by 
  Mid-career Researcher Program through NRF grant funded by the MEST (No.2010-0013875).
S.C.K., J.K., J.H.L., and C.H.R. are
  members of the Dedicated Researchers for Extragalactic AstronoMy (DREAM) team 
  in the Korea Astronomy and Space Science Institute (KASI).
This paper is based on observations made with in the NASA/ESA Hubble Space Telescope,
  obtained from the data archive at the Space Telescope Science Institute, which is
  operated by AURA, Inc., for NASA under contract NAS 5-26555. 
This research has made use of NASA's Astrophysics Data System Abstract Service
  and the NASA/IPAC Extragalactic Database (NED), which is operated by the
  Jet Propulsion Laboratory, California Institute of Technology, under contract
  with the National Aeronautics and Space Administration.


\begin{table*}[t]  
\begin{center}
{\bf Table 1.}~~Basic Information of NGC 1156 \\
\vskip 3mm
{\small 
\setlength{\tabcolsep}{1.2mm}
\begin{tabular}{lccl} \hline\hline  
Parameter & Information & Reference \\
\hline 
$\alpha_{J2000.0}$, $\delta_{J2000.0}$ & 02$^h$ 59$^m$ 42.19$^s$, $+25\arcdeg$ $14\arcmin$ 14.2\arcsec & NASA/IPAC Extragalactic Database (NED) \\
$l, b$ & 156.\arcdeg31, $-29.\arcdeg20$  & NED \\
Morphological type     & IB(s)m V-VI     & \citet{devaucouleurs91} \\
Position angle (N through E) & 39\arcdeg    & \citet{hunter02} \\
Angular diameter, $D_0$      & $3.3\arcmin$ & \citet{devaucouleurs91} \\
Axial ratio                  & 7.4          & \citet{devaucouleurs91} \\
Kinematical axes of the ionized gas, & 84\arcdeg & \citet{hunter02} \\
   \hskip 9mm neutral gas, and stellar disks    &   & \\
Inclination, $i$       & 42\arcdeg              & \citet{bottinelli84} \\
Radial velocity, $v_r$ & $375 \pm 1$ \kms & NED \\
Redshift, $z$          & 0.001251 & NED \\
Distance modulus, $(m-M)_0$ & $29.02\pm0.40$ mag ($d = 6.4 \pm 1.2$ Mpc)& \citet{tully88} \\
                    & $29.46\pm0.15$ mag ($d = 7.8 \pm 0.5$ Mpc)& \citet{karachentsev96} \\
                    & $29.39\pm0.20$ mag ($d = 7.6 \pm 0.7$ Mpc)& This study \\
Reddening, $E(B-V)$    & $0.35 \pm 0.05$ mag             & This study \\
$B_0$, $M_{B,0}$       & $11.78 \pm 0.10$ mag$^{\rm a}$, $-17.68$ mag$^{\rm b}$ & \citet{barazza01} \\
$B_0$                  & $11.56$ mag$^{\rm c}$           & \citet{tully88} \\
$B_0$                  & $11.61$ mag$^{\rm a}$           & \citet{bottinelli84} \\
$B_0$, $M_{B,0}$       & $12.32$ mag, $-17.85$ mag$^{\rm b,d}$ & \citet{devaucouleurs91} \\
$V_0$          & $11.31$ mag$^{\rm a}$           & \citet{barazza01} \\
$M_V$                  & $-18.67$ mag$^{\rm b}$  & \citet{hunter04} \\
$R_0$          & $10.91$ mag$^{\rm a}$           & \citet{barazza01} \\
$R$                    & $11.91 \pm 0.04$ mag    & \citet{james04} \\
$(B-V)_0$              & $0.47^{\rm a}$        & \citet{barazza01} \\
$(B-R)_0$              & $0.87^{\rm a}$        & \citet{barazza01} \\
Effective radius, $r_{\rm eff}^B$, $r_{\rm eff}^V$, $r_{\rm eff}^R$ 
   & $31.58\arcsec, 34.28\arcsec, 35.48\arcsec$ 
   & \citet{barazza01} \\
Effective surface brightness, & $21.28, 20.98, 20.66$ mag arcsec$^{-2}$ & \citet{barazza01} \\
  \hskip 9mm $\langle \mu \rangle_{\rm eff}^B$, $\langle\mu\rangle_{\rm eff}^V$, 
  $\langle\mu\rangle_{\rm eff}^R$ & & \\
flux density at $12, 25, 60, 100 \micron$   & 0.17, 0.55, 5.24, 10.48 Jy & \citet{dale00} \\
flux density at $6.75, 15 \micron$ &$0.09\pm0.02$, $0.14\pm0.03$ Jy & \citet{dale00} \\
\hi flux               & 71.3 Jy \kms & \citet{swaters02} \\
                       & 72.72 Jy \kms & \citet{haynes98} \\
                       &$75.6\pm6.4$ Jy \kms & \citet{minchin10} \\
\hi mass, $M_{H_I}$      & $1.02 \times 10^9$ \Msun & \citet{swaters02} \\
                       & $(1.08\pm0.09) \times 10^9$ \Msun & \citet{minchin10} \\
\hi line width$^{\rm e}$, $W_{50}$ & $73 \pm 3$ \kms & \citet{broeils94} \\
$M_{H_I}/L_B$            & $0.56$ \Msun/$L_\odot$  & \citet{swaters02} \\
Star formation rate, SFR & $0.71 \pm 0.07$ \Msun/yr & \citet{james04} \\
\hline 
\end{tabular}
} 
\end{center}
{\small
\hskip 6mm $^{\rm a}$ Only the foreground extinction is corrected

\hskip 6mm $^{\rm b}$ Assuming d $= 7.8 \pm 0.5$ Mpc

\hskip 6mm $^{\rm c}$ Both foreground and internal extinctions are corrected

\hskip 6mm $^{\rm d}$ Assuming $A_B = 0.71$ mag 

\hskip 6mm $^{\rm e}$ Profile width at a level of 50\% of the peak value, corrected
  for instrumental broadening
}
\end{table*}

\begin{table*}[t]
\begin{center}
{\bf Table 2.}~~Selected Blue and Red Supergiant Stars in NGC 1156$^{\rm a}$ \\
\vskip 3mm
{\small
\begin{tabular}{ccccccccccc} \hline\hline
ID & R.A.(J2000.0)$^{\rm b}$ & Decl.(J2000.0)$^{\rm c}$ & $U$ & $Uerr$ & $B$ & $Berr$ & $V$ & $Verr$ & $I$ & $Ierr$ \\
\hline
BSG1 & 2:59:41.96 & 25:14:33.6 & 21.130 & 0.017 & 21.598 & 0.013 & 21.263 & 0.014 & 20.827 & 0.015 \\
BSG2 & 2:59:42.02 & 25:14:08.3 & 21.786 & 0.026 & 22.009 & 0.017 & 21.772 & 0.019 & 21.453 & 0.022 \\
BSG3 & 2:59:41.48 & 25:14:13.8 & 21.834 & 0.028 & 22.221 & 0.022 & 22.004 & 0.025 & 21.584 & 0.028 \\
\\
RSG1 & 2:59:42.38 & 25:14:33.7 & 26.942 & 3.690 & 25.184 & 0.181 & 22.674 & 0.030 & 20.617 & 0.013 \\
RSG2 & 2:59:42.12 & 25:14:27.9 & 25.513 & 0.607 & 24.757 & 0.110 & 22.708 & 0.031 & 21.199 & 0.019 \\
RSG3 & 2:59:42.57 & 25:14:34.3 & 25.554 & 0.788 & 25.305 & 0.177 & 22.893 & 0.036 & 20.796 & 0.014 \\
\hline
\end{tabular}
} 
\end{center}
\hskip 6mm $^{\rm a}$ BSGs are in the order of $B$-magnitudes and 
  RSGs are in the order of $V$-magnitudes.

\hskip 6mm $^{\rm b}$ in hours, minutes, and seconds

\hskip 6mm $^{\rm c}$ in degrees, arcminutes, and arcseconds
\end{table*}

\end{document}